\documentclass{sig-alternate-05-2015}
\usepackage{enumitem}

\begin{document}

\setcopyright{acmcopyright}

\doi{10.475/123_4}

\isbn{123-4567-24-567/08/06}

\conferenceinfo{PLDI '13}{June 16--19, 2013, Seattle, WA, USA}

\acmPrice{\$15.00}

%
\conferenceinfo{WOODSTOCK}{'97 El Paso, Texas USA}

\title{High-Performance Ultrasonic Levitation with FPGA-based Phased Arrays} 
%
%
%
%
%

\numberofauthors{6} 
%
\author{
\alignauthor
William Beasley\\
       \affaddr{Department of Electrical and Electronic Engineering}\\
       \affaddr{Woodland Road}\\
       \affaddr{University of Bristol, UK}\\
\alignauthor
Brenda Gatusch\\
       \affaddr{Department of Electrical and Electronic Engineering}\\
       \affaddr{Woodland Road}\\
       \affaddr{University of Bristol, UK}\\
\alignauthor
Daniel Connolly-Taylor\\
       \affaddr{Department of Electrical and Electronic Engineering}\\
       \affaddr{Woodland Road}\\
       \affaddr{University of Bristol, UK}\\
\and 
\alignauthor 
Chenyuan Teng\\
       \affaddr{Department of Electrical and Electronic Engineering}\\
       \affaddr{Woodland Road}\\
       \affaddr{University of Bristol, UK}\\
\alignauthor 
Asier Marzo\\
       \affaddr{Department of Electrical and Electronic Engineering}\\
       \affaddr{Woodland Road}\\
       \affaddr{University of Bristol, UK}\\
\alignauthor 
Jose Nunez-Yanez\\
       \affaddr{Department of Electrical and Electronic Engineering}\\
       \affaddr{Woodland Road}\\
       \affaddr{University of Bristol, UK}\\
}
\additionalauthors{Additional authors: John Smith (The Th{\o}rv{\"a}ld Group,
email: {\texttt{jsmith@affiliation.org}}) and Julius P.~Kumquat
(The Kumquat Consortium, email: {\texttt{jpkumquat@consortium.net}}).}
\date{30 July 1999}

\maketitle
\begin{abstract}
We present a flexible and self-contained platform for acoustic levitation research based on the Xilinx Zynq SoC using an array of ultrasonic emitters. The platform employs an inexpensive ZedBoard and provides fast movement of the levitated objects as well as object detection based on the produced echo. Several features available in the Zynq device are of benefit for this platform: hardware acceleration for the phase calculations, large number of parallel I/Os connected through the FPGA Mezzanine connector (FMC), integrated ADC capabilities to capture echo signals and ease of programmability due to a C-based design flow for both CPU and FPGA. A planar and spherical cap phased arrays are created and we investigate the capabilities and limitations of the different designs to improve the stability of the levitation process.  
\end{abstract}

%
%

%
%

%
%
\printccsdesc



\section{Introduction}
Acoustic levitation employs radiation forces to suspend an object in air without any contact. This type of levitation has been widely used for contactless manipulation of particles in biology, medicine, and physics, amongst other areas. Other methods exist for contactless handling, including magnetic, electrostatic and optical trapping. However, acoustic processing has the advantage of working on different materials and at different sizes. Therefore it can manipulate insulators or conductors and magnetic or non-magnetic objects with acoustic forces. Typical applications range from cell manipulation, blood washing for lipids separation, and sample isolation before x-ray radiation of proteins, to crystallographic experiments and manufacturing of small electronic devices. Acoustic levitation can be generated with an ultrasonic phased array. This device consists of a group of transducer elements emitting with the sample amplitude and frequency but different phases. The waves emitted by element interfere with each other forming a beam pattern. Focusing of sound along a desired direction is possible with millimetric accuracy and within milliseconds. In this paper we investigate the suitability of a FPGA-based device to create a high-performance and low-cost levitation device capable of working in a closed loop configuration and receive an echo signal from the levitating objects. We focus on exploiting the capabilities of the Zynq SoC including relative low-cost, ease of programmability and user interaction, extended parallel I/O capabilities, high-performance via hardware acceleration and mix-signal capabilities with integrated ADC conversion. The overall aim is to use these Zynq features to construct a low-cost alternative to high-end commercial platforms suitable for research. The contributions of this work are:\begin{enumerate}[topsep=0pt,itemsep=-1ex,partopsep=1ex,parsep=1ex] \item We demonstrate the suitability of the Zynq SoC device at building and controlling different types of ultrasonic systems including flat and spherical cap systems.\item We demonstrate the positive levitation characteristics achieved with the hardware acceleration.\item We show a viable closed loop system that can be used to not only levitate objects but also to detect the presence of objects and perform position corrections.\item We construct an open-source system including PCB schematics, software and hardware to encourage further research in the area. \end{enumerate} 

\section{RELATED WORK}
There is a large range of applications that can use ultrasonic phased arrays and this has resulted in research both at the commercial and academic level. For example, microfluidic separation chips, also referred to as Lab-on-a-Chip, can make use of ultrasonic transducers placed underneath them to generate a standing wave perpendicularly to the flow channel to separate particles [1]. Other innovative uses are haptic technologies with ultrasonic waves used to recreate tactile sensations in mid-air [2]. A mid-air tactile display, called the Airborne Ultrasound Tactile Display (AUTD), has been created to produce tactile stimulation from a distance. The display makes use of airborne ultrasound to stimulate human skin. It uses the phased array focusing technique to produce radiation pressure that can press the human skin in the direction of propagation. This technology is expected to be augmented with tactile feedback floating images displays and other user interfaces [3]. The application of FPGAs to this area has been studied in [4] that presents a reconfigurable, cost-effective FPGA and PC-based ultrasonic system, designed for teaching and medical imaging research. The system uses the MD2131 beamformer source driver from Microchip Technology Inc. to generate arbitrary waveforms and an analogue front-end to obtain maximum flexibility and data access to various ultrasonic data streams. The results obtained from applications such as plane wave excitation and delay-and-sum beamforming show that the open platform can help biomedical students and researchers to develop and evaluate different imaging strategies for medical ultrasonic imaging and non-destructive testing techniques. Most of the processing is done on the host PC and the total cost of the system is \$26K which is much higher than the cost of the proposed system. A comparison of FPGAs versus other technologies like DSPs and microcontrollers is addressed in [5]. The FPGA is a Virtex-5 device but the ultrasonic signal processing algorithms such as FFTs are not implemented with custom logic but with a Microblaze 32-bit soft-core RISC processor directly programmed in C which results in poor performance. The platform deals exclusively with non-destructive testing using ultrasonic signals and no levitation applications are proposed for the system. The same team extends the previous work to a Zynq device in [6] and in this case the Microblaze is replaced by the ARM processor which achieves a 30x speed-up in the  computing the ultrasonic processing algorithms. The FPGA programmable logic is not used and the solutions are based on the available Cortex A9 devices.  The work in [7] also discusses how an FPGA can be used to control the generation of signals for the actuators power drivers but the system is based on a vibration ring and not on a phased array. No information is available on how the FPGA card is used but most of the processing takes place in a host PC. In contrast our solution is self-contained and all the required computation is performed inside the FPGA device. Other commercial examples include [8] that creates an electromechanical system with a fast, adaptive digital control to allow near-field acoustic levitation based on boards that integrate Kintex FPGAs. By controlling the actuator vibration pattern of standing and travelling waves, a levitated object is manipulated by air pressure only. A resonance tracking system is employed to keep the system within its resonance bandwidth. No details on the function that the FPGAs perform are available. Ultraino is a phased-array system based on an Arduino Mega [9]. It supports emission on 64 channels at 40kHz, however due to the limitations of the Arduino MEGA only 10 divisions per period are achieved and the update rate is below 100Hz.
\section{Ultrasonic levitation platform description}
\begin{figure*}[!htbp]
\centering
\includegraphics[width=14cm]{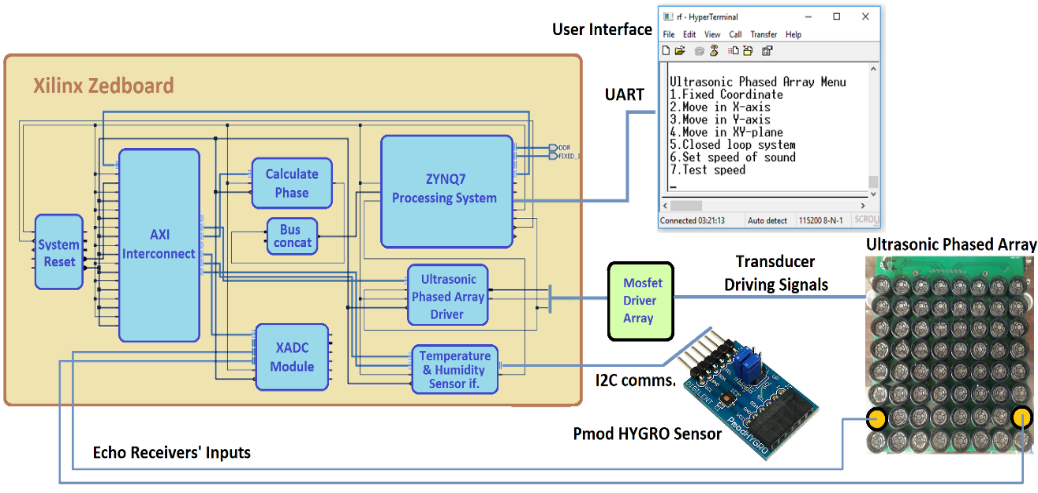}
\caption{Overview of full system}
\label{fig:sys}
\end{figure*}
Figures ~\ref{fig:sys} shows a full overview of the system with an example of a flat UPA (ultrasonic phased array) developed in-house. The platform is based on a Zedboard that interfaces the ultrasonic phased array using a Samtec FMC connector that exposes 68 single-ended I/O. The array uses 64 channels (8x8) and two of these channels can be connected to the two channel ADC connector available in the Zynq device. These two channels can be used to receive ultrasonic signals while the rest are used in emitter mode. The processing system in the Zynq device runs the software that controls the phased array driver and calls a hardware function called phase\_calculate to perform the focal point calculations. It also offers a simple interface so that the user can interact with the system. Using this interface the user can specify the desired focal point as a set of 3d coordinates in the space above the array. The processing system facilitates the calculation of the required set of phase delays to produce a focal point at a user defined location.The phase delays, expressed as the number of clock cycles to delay the signal of a channel, are then written directly into the Ultrasonic Phased Array Controller (UPAC) registers over the AXI bus. The UPAC then generates the required 40kHz signals and outputs them to the FMC port. The phase shifted 40kHz signals are then amplified from their 0-3.3v range to 0-16v using MOSFET drivers (TC4427a) and the amplified signals are then fed to their respective transducers. The following sections describe in more detail the functions in Figure ~\ref{fig:sys}. 

\subsection{Ultrasonic Phased Array}

The Zedboard is limited to a maximum voltage and current rating of 3.3V at 2A[1]. This voltage is not sufficient for driving the transducers, hence each channel needs to be amplified. To achieve this, MOSFET drivers are used to step up the control signal voltage from the 0V-3.3V range to the 0-16V range. The TC4427A MOSFET driver from Microchip is used as it combines two MOSFET drivers in a single SOIC-8 package.  The chosen array size needs 64 IOs which means that the single FMC connector available in the Zedboard is sufficient. For conversion between electrical and mechanical energy electro-acoustic transducers are used that utilise the Reverse Piezoelectric effect by applying a potential difference across a piezo crystal. The transducer array is composed of MCUSD16P40B12RO or 250ST180 transducers that have an operating frequency of 40 kHz and 25 kHz respectively. From a practical point of view the transducers are very effective at attenuating signals outside of their operating range. Therefore, it is not essential to drive them with a sinusoid. It is simpler to avoid additional hardware complexities and drive them with a square wave generated in the FPGA logic. In a flat system the drivers and transducers can be combined into one PCB that provides the physical mounting point as well as routing for all control signals. The spherical cap system needs a physical separation between the transducers and the driver electronics.

\subsection{Calculating the Phases}


If r\textsubscript{i}(x,y,z) represents the three dimensional Cartesian coordinates of the ith transducer within the array and r\textsubscript{f}(x,y,z) represents the same values for target focal point respectively, the path length LP\textsubscript{i} between them can be expressed as the Euclidean distance as shown in Equation~\ref{fig:eq1}. The required phase shift for the i\textsubscript{th} transducer, in wavelengths, is the remainder after dividing the path length by the wavelength. Then, this can be multiplied by $2\pi$ to obtain the phase shift in radians as shown in Equation~\ref{fig:eq2}. 
For phased arrays, the size of the focal point (w) can be expressed as a function of wavelength ($\lambda$), focal length (R) and the side length of the array (D) as shown in Equation~\ref{fig:eq3}. In general, for acoustic levitation the levitated particles have to be smaller than half-wavelength, in our case that is 4mm approximately.

\begin{equation} \label{fig:eq1}
\includegraphics[width=8cm]{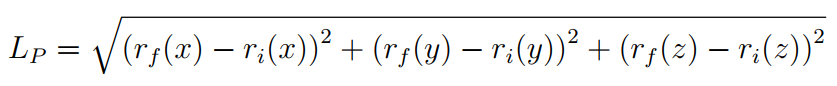}
\end{equation}

\begin{equation} \label{fig:eq2}
\includegraphics[width=4cm]{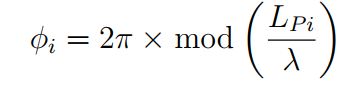}
\end{equation}


\begin{equation} \label{fig:eq3}
\includegraphics[width=2cm]{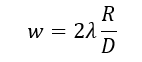}
\end{equation}

\begin{figure}[h]
\centering
\includegraphics[width=9cm]{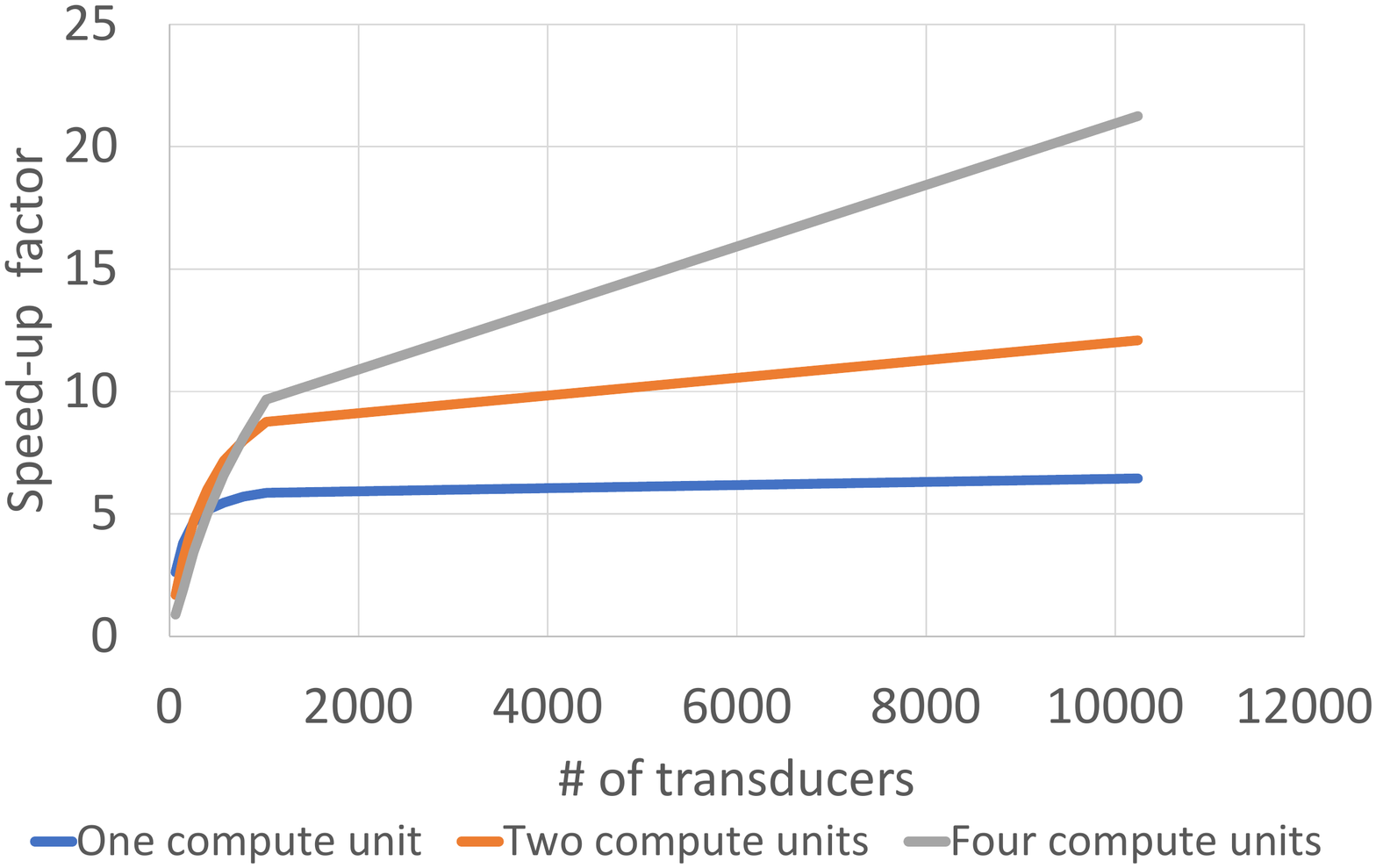}
\caption{Calculate phase Hardware acceleration}
\label{fig:pc}
\end{figure}
\begin{figure}[h]
\centering
\includegraphics[width=9cm]{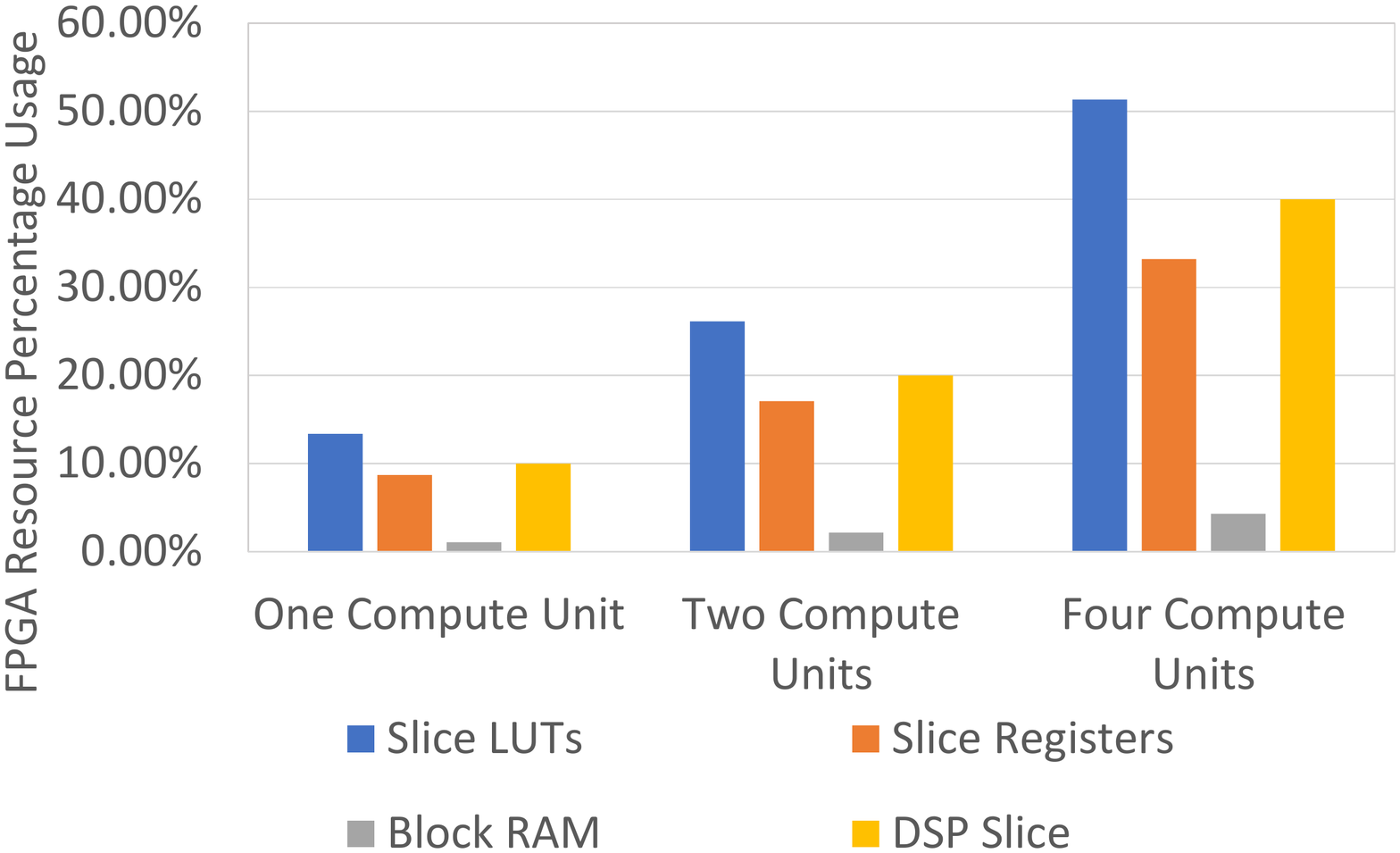}
\caption{Calculate phase Hardware Resources}
\label{fig:res}
\end{figure}

When the acoustic pressure waves arrive concurrently at a point, they will interfere with each other creating an oscillating acoustic field. Within this acoustic field, particles will experience a non-linear acoustic effect called the Acoustic Radiation Force (ARF). The ARF has a non-zero time-average sum, and as such a static pressure is applied to the particle surface. Performance is important when calculating the phase-shifts specially if the system needs to maintain more than one focal point in parallel. Typically, for research applications, it is necessary that the array produces multiple focal points in parallel. In our work, multiple focal points can be created by multiplexing focal points in time. In this time division multiplexing, the array only produces one focal point at any instance in time. However, by cycling though the desired number of focal points, with suitable speed, the effect of multiple focal is created. To speed up the phase calculations an accelerator using the SDSoC high-level synthesis tools has been created. SDSoC uses Vivado HLS compilers to generate the hardware implementation and it also creates the drivers needed to communicate the accelerator implemented in the FPGA with the ARM host in the Zynq board. Pragmas in the source code can be used to control the type of data movement cores that are implemented and it also possible to use pragmas to change the number of compute units that are implemented in parallel. The accelerator block takes as inputs the target three dimensional coordinates and an offset identifying the active transducer. It implements the equations shown in ~\ref{fig:eq1} and ~\ref{fig:eq2}. A number of optimisations have been applied to unroll and pipeline the for loop that accumulates four partial results. The output of the accelerator is the phase delay for the corresponding transducer. Figure~\ref{fig:pc} measures the performance speed up of the FPGA compute units compared with the software versions running on the ARM core. One compute unit with a 64 transducer frame achieves a speed-up factor of aproximately 2.7. By accelerating the phase delay calculation with one compute unit, the system latency decreases from 154us to 60us. This reduction led to an increase in refresh rate from 6.49 kHz to 16.6 kHz. Overall, increasing the number of compute units to more than one is only helpful when the number of computed phase delays increases. This will be possible with set-ups with more than 64 transducers or if a single call was used to calculate many transducer frames in parallel. For example the configuration with 4 compute units achieves an speed-up factor of 21 processing a batch of 160 frames with 64 transducer each. The complexity of the design with different number of compute units is shown in Figure~\ref{fig:res} and the implementation clock frequency is 100 MHz for the FPGA and 666 MHz for the Cortex A9 processor. To measure power we have implemented the hardware on a zc702 board that enables direct power measurements of the voltage regulators that supply power to the processor and the FPGA components using the power manager BUS (PMBUS). The measure total power on the Cortex A9 processor VCCPINT is approximately 700 mW while the corresponding power on the FPGA device is 520 mW in the VCCINT rail. As expected the configurable accelerator is more power efficient and combining this lower power with the time speed-up the overall result will be a much more energy efficient solution than the programmable processor.

\subsection{HYGRO sensor}

\begin{figure}[!htbp]
\centering
\includegraphics[width=8cm]{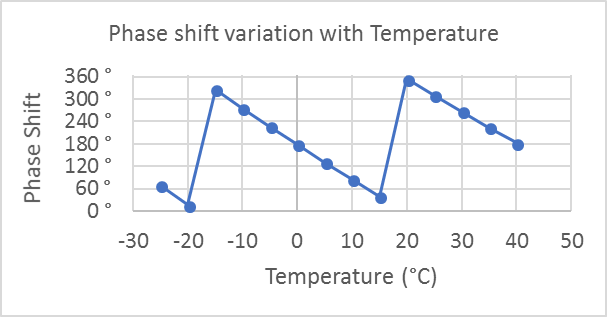}
\caption{Temperature effects on phase}
\label{fig:temp1}
\end{figure}

\begin{figure}[!htbp]
\centering
\includegraphics[width=8cm]{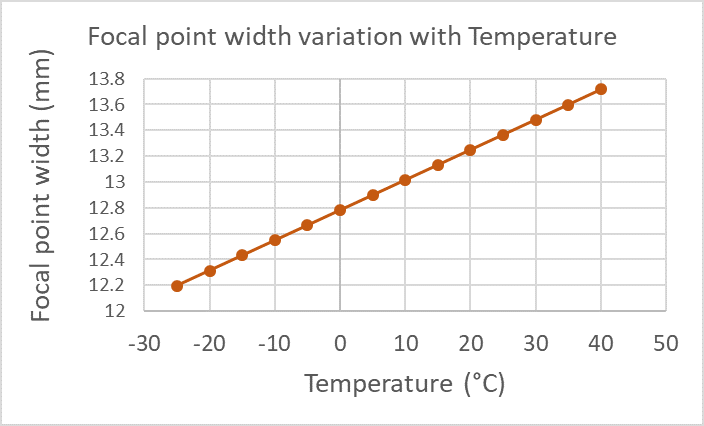}
\caption{Temperature effects on focal point size}
\label{fig:temp2}
\end{figure}

During the process of testing single-beam levitation of a particle, it was observed that the particle stability and ability to levitate in mid-air were significantly affected by the characteristics of the environment, mainly temperature variations. The reason is that the speed of sound increases linearly with temperature and this affects the propagation of the acoustic waves and how the focal point is formed. If the speed of sound increases, the wavelength of the sound wave also becomes larger with temperature. Considering that the wavelength value is used to calculate the phase shifts of the transducer driving signals, this means that the phase shifts are also affected. Since the wavelength is also used to determine the width of the focal point by $w=2\lambda R/D$, this measurement will also change. Figure~\ref{fig:temp1} and Figure~\ref{fig:temp2} shows the phase delay and focal size variation due to temperature in the 8x8 phased array when the focal length R is 100 mm.To compensate for temperature variations affecting the speed of sound, a Pmod HYGRO sensor was added to the system. This sensor can digitally report the ambient temperature upon request by the host board with up to 14 bits of resolution. It provides a temperature accuracy of $±0.2 °C$. Communication with this board is performed via a 6-pin Pmod connector with I2C interface. The processing system reads a temperature measurement when requested by the user, and updates the speed of sound value and its wavelength. 

\section{Developed Phased Array types}

\begin{figure}[!htbp]
\centering
\includegraphics[width=7cm]{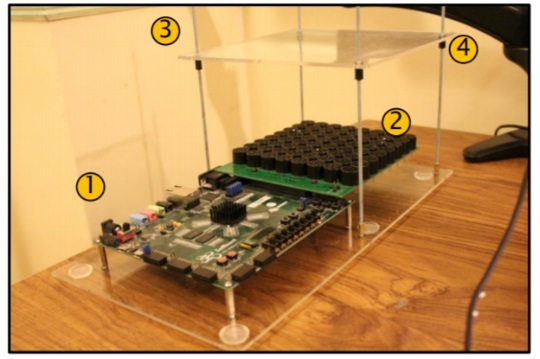}
\caption{Flat phase array configuration}
\label{fig:flat}
\end{figure}

The array elements can be arranged in different ways to achieve the acoustic pressure necessary to overcome gravity and suspend an object in the air. In this work, to investigate different degrees of movement and stability during particle control, flat single-sided and reflector-plate arrays have been considered, as well as phased arrays based on spherical cap single-sided and  double-sided. The flat single-sided with reflector plate is shown in Figure~\ref{fig:flat}, it has the advantages of being easy to construct and able to produce different types of traps such as twin-trap, vortex trap or bottle trap.  

\begin{figure}[!htbp]
\centering
\includegraphics[width=7cm]{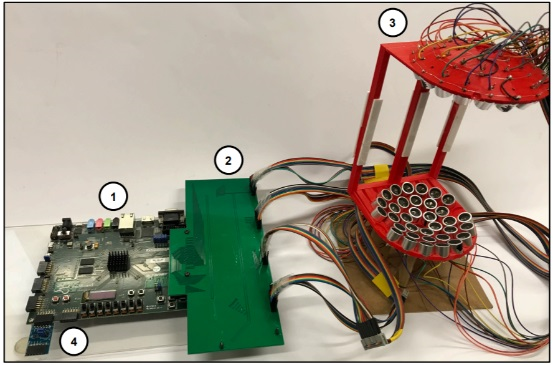}
\caption{spherical cap phase array configuration}
\label{fig:spherical cap}
\end{figure}

There are limitations in terms of stability when the reflector plate is removed due to the lack of acoustic pressure that can be supplied to the focal point by a single-beam. The reflector plate creates a standing wave and improves stability considerably. The incoming and reflected wave interact with each other and when a compression meets another compression or a rarefaction meets another rarefaction, amplitude doubles. The areas of minimum pressure in a standing wave are called nodes. And those of maximum pressure are called antinodes. Floating particles will be suspended just below the nodes. The spherical cap array shown in Figure~\ref{fig:spherical cap} provides a more stable trap than the flat array. It requires the same type of phase delay pattern as the flat array but stability increases due to the formation of a standing wave.

\section{Open-loop levitation test}

\begin{figure}[!htbp]
\centering
\includegraphics[width=7cm]{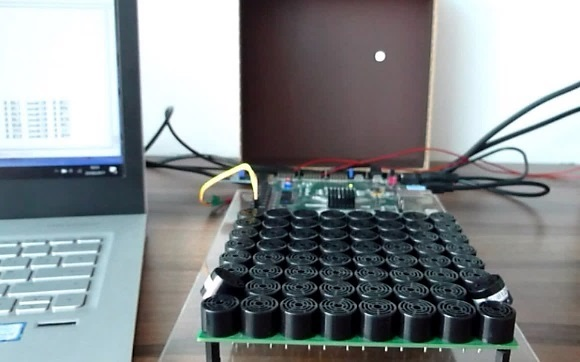}
\caption{Flat array Levitation test}
\label{fig:lev1}
\end{figure}

\begin{figure}[!htbp]
\centering
\includegraphics[width=7cm]{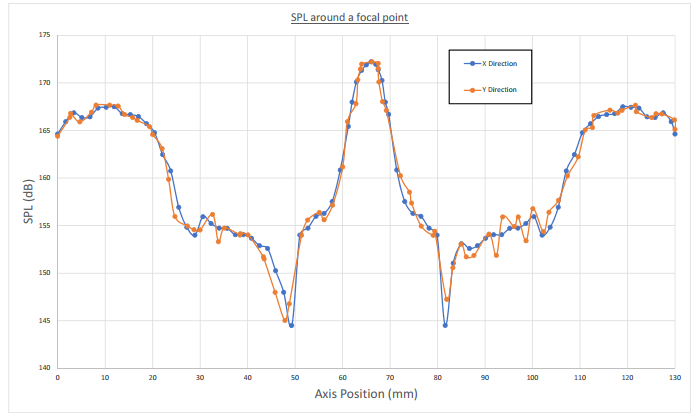}
\caption{Focal point strength analysis}
\label{fig:spl}
\end{figure}

\begin{figure}[!htbp]
\centering
\includegraphics[width=9cm]{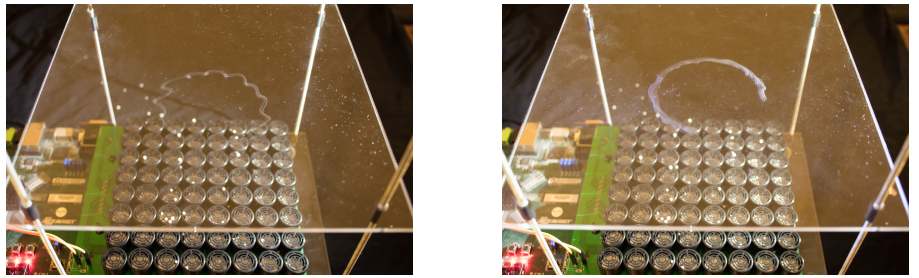}
\caption{Particle control comparison with software (left) and hardware (right) phased calculations. }
\label{fig:error}
\end{figure}

Figure~\ref{fig:lev1} demonstrates successful particle levitation by the system. Indicators of the quality of the levitation system include strength, stability and refresh rate. To quantify the intensity or strength of a focal point the sound-pressure-level (SPL) can be measured. To measure the SPL, an ultrasonic transducer, operating as a microphone, was placed at coordinates (66,66,100cm) and the array was instructed to produce a focal point at this position. The terminals of the microphone were subsequently connected to an oscilloscope, to measure the Vrms signal generated by the transducer. The manufacturer provides a sensitivity value that can be used to calculate the SPL for a given Vrms. To obtain a profile of the SPL across a focal point, the microphone was mounted on a set of sliding arms, that could be used to adjust its x and y coordinates. The profile for the SPL can be seen in Figure~\ref{fig:spl}. The selectivity of the SPL response can be predicted to conform to the focal size give in  equation ~\ref{fig:eq2}. Using this equation with $\lambda\approx8.5mm$, focal length R = 100mm, and side length D = 132mm the focal point should be $\approx13mm$ in width. In the experiment, a peak SPL of 172dB was recorded, with a focal point width of approximately 13mm, which aligns with the expected outcomes. The refresh rate is analogous to video processing systems, in which the performance is measured in frame rate (e.g 60fps). In this case the refresh rate (Hz) defines the temporal resolution between successive phase patterns. For example, a refresh rate of 2KHz gives a temporal resolution of 0.5ms. The maximum refresh rate is therefore limited by the time taken to calculate and generate a new phase pattern for all the transducers. A higher refresh rate should yield better performance in the movement of particles, as it allows to use a smaller step size for a given speed. The tests were performed with a 1 mm diameter expanded polystyrene sphere (EPS) with a density of 29.63 kg/m3 (Custompac Ltd.,Castleford, UK). To measure linear speeds, the EPS particle was placed in the center of the array at a height of 100 mm. The particle was then moved forwards and backwards along a linear path of 10 mm. Upon successful completion of 10 iterations, the speed was increased. The software-only and accelerated implementations operated at their maximum refresh rate, and slowly incremented the step size. The software implementation required a much larger step size to obtain the same speed as the hardware accelerated implementation. This confirms that the increased refresh rate enables a finer control of the levitated particles. The absolute maximum speed was assessed by moving the particle in a continuous circular orbit of radius 30mm around the centre of the array. Upon the completion of 5 orbits, the speed was increased until the particle was ejected out of the array. Table ~\ref{table:speed} shows that the EPS spheres can travel at much higher speeds when moved around a circle. The EPS particles could reach speeds of up to 460 mm/s, which is a 14.8\% over the maximum stable speed (392 mm/s). Once again it should be noted that the software and hardware methods achieved a similar maximum speed however the step size of the software method needs to be considerably larger. This has a negative effect on the overall stability of the system as shown in Figure ~\ref{fig:error} that shows the error in the orbit using long exposure photographs.

\begin{table}
\caption{Particle speed analysis}
\label{table:speed}
\resizebox{\columnwidth}{!}{
\begin{tabular}{c*{4}{c}}
    & \text{Speed (mm/s)} & \text{Step size (mm)}  & \shortstack{Normalized \\ speed (mm/s)} & \shortstack{Fixed step \\ size (mm)} \\
\hline
\shortstack{Linear \\ software}  & 385 & 0.05929 & 168 & 0.026 \\
\hline
\shortstack{Linear \\hardware}  & 392 & 0.026 & 392 & 0.026 \\
\hline
\shortstack{Circular \\software}  & 450 & 0.0709 & 197 & 0.0304 \\
\hline
\shortstack{Circular \\ hardware}  & 460 & 0.0304 & 460 & 0.0304 \\
\end{tabular}
}
\end{table}

\section{Close-loop echo test}

\begin{figure}[!htbp]
\centering
\includegraphics[width=7cm]{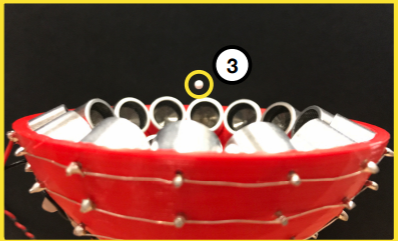}
\caption{spherical cap levitation test}
\label{fig:lev2}
\end{figure}

\begin{figure}[!htbp]
\centering
\includegraphics[width=7cm]{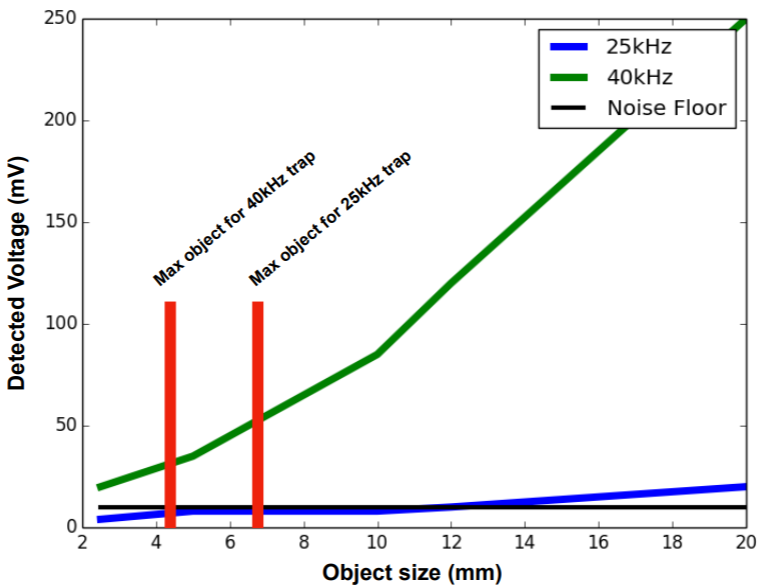}
\caption{Transducer frequencies and echo signal}
\label{fig:echo2}
\end{figure}

\begin{figure}[!htbp]
\centering
\includegraphics[width=7cm]{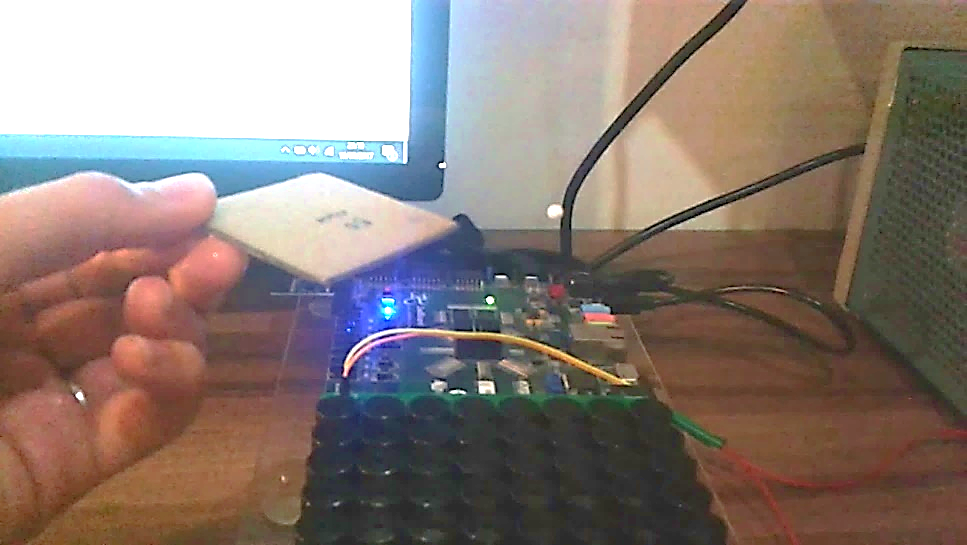}
\caption{Object detection in Flat Array}
\label{fig:echo}
\end{figure}

To create the close-loop system two transducers are used as receivers instead of transmitters. The system is  limited to two detecting transducers since the the 12-bit ADC available in the Zedboard has 2 channels that  operate at up to 1 MSPS. Initial echo tests with the flat array (Figure ~\ref{fig:echo}) indicated that using the same frequency for levitation and object detection (i.e. 40 kHz) was not feasible. To address this issue, two transducers were replaced with transducers operating at a different frequency. A new arrangement was designed. The transducers that generate levitation operate at 25 kHz and can be arranged in a flat or spherical cap configuration with 64 transducers in total. In the spherical cap configuration, 32 transducers were placed on the top and 32 on the bottom, two 40kHz transducers were added to detect the levitated particle. A particle levitating with the spherical cap array can be seen in Fig ~\ref{fig:lev2}. Fig ~\ref{fig:echo2} shows the echo signal received in the detecting transducers depending on the frequency of the transducers. The X axis indicates the size of the levitated particle and the Y axis the amplitude of the echo signal. The blue line shows the result when the 25 kHz transducers are used for detection with 40 kHz being used for levitation, and the green line the opposite. It is clear that the green line represents a stronger signal while it also has the advantage that traps generated with 25 kHz can lift larger particles. The red columns indicate the maximum particle size depending on the frequency. This setup enables to detect the position of the particle in 4 directions west-east-north-south. We are working on a new set up that aims at increasing the echo resolution with more transducers and by using additional signal processing to be able to build an accurate echo image from the particle. 

\section{Conclusions and future work}

This paper has shown a high-performance implementation of an ultrasonic levitating system using a low-cost and integrated solution based on a Xilinx Zynq device. Different types of phased array configurations and device features are used to enable stable and accurate particle control in open-loop and closed-loop configurations. The paper shows that the high-performance Zynq SoC offers a feature-rich and low-cost platform for this application. The following links show different demonstration videos with a circular pattern \verb|https://youtu.be/GP1x6JX_aTI|, echo detection \verb|https://youtu.be/4r_L2R9H_Ws|,  XY plane movement \verb|https://youtu.be/q1u2oCEx3Nk| and speed \linebreak \verb|https://youtu.be/pNvneg7r7fM|. The following github link contains the necessary materials to recreate the project using a xilinx Zedboard https://github.com/eejlny/ultrasonic-levitation-with-Xilinx-Zynq including source code and schematics. Future work involves improving the echo mechanism with more receiving sensors and multiple particule levitation. The high frame rates possible with multiple compute units will enable the levitation of multiple particles in parallel with fast multiplexation of the acoustic field. This would enable novel applications to create graphic representations in human-machine iteration. 

\section*{Acknowledgments}

This work was partially supported by Xilinx and UK EPSRC with the ENPOWER (EP/L00321X/1) and the ENEAC \linebreak (EP/N002539/1) projects.

\addtolength{\textheight}{-12cm}   




\end{document}